\documentclass[twocolumn, aps, superscriptaddress]{revtex4}

\usepackage{graphicx}
\usepackage{amsmath,amssymb}
\usepackage{hyperref}
\hypersetup{colorlinks=true,linkcolor=blue,urlcolor=blue,citecolor=red}

\usepackage[normalem]{ulem}
 \usepackage{natbib}

\usepackage[utf8]{inputenc}
\usepackage{array}
\usepackage{diagbox}
\usepackage{caption}
\usepackage{mathtools}

\begin{document}

\title{Ground motion prediction at gravitational wave observatories using archival seismic data}

\author{Nikhil Mukund}
\affiliation{Inter-University Centre for Astronomy and Astrophysics (IUCAA), Post Bag 4, Ganeshkhind,  Pune 411 007, India}

\author{Michael Coughlin}
\affiliation{Division of Physics, Math, and Astronomy, California Institute of Technology, Pasadena, CA 91125, USA}

\author{Jan Harms}
\affiliation{Gran Sasso Science Institute (GSSI), I-67100 L'Aquila, Italy \\
INFN, Laboratori Nazionali del Gran Sasso, I-67100 Assergi, Italy}

\author{Sebastien Biscans}
\affiliation{LIGO, California Institute of Technology, Pasadena, CA 91125, USA}
\affiliation{LIGO, Massachusetts Institute of Technology, Cambridge, MA 02139, USA}

\author{Jim Warner}
\affiliation{LIGO Hanford Observatory, Richland, WA 99352, USA}

\author{Arnaud Pele}
\affiliation{LIGO Livingston Observatory, Livingston, LA 70754, USA}

\author{Keith Thorne}
\affiliation{LIGO Livingston Observatory, Livingston, LA 70754, USA}

\author{David Barker}
\affiliation{LIGO Hanford Observatory, Richland, WA 99352, USA}

\author{Nicolas Arnaud}
\affiliation{LAL, Univ. Paris-Sud, CNRS/IN2P3, Universit\'e Paris-Saclay, F-91898 Orsay, France}
\affiliation{European Gravitational Observatory (EGO), I-56021 Cascina, Pisa, Italy}

\author{Fred Donovan}
\affiliation{LIGO Laboratory, Massachusetts Institute of Technology, Cambridge, MA 02138, USA}

\author{Irene Fiori}
\affiliation{European Gravitational Observatory (EGO), I-56021 Cascina, Pisa, Italy}

\author{Hunter Gabbard}
\affiliation{SUPA, School of Physics and Astronomy, University of Glasgow, Glasgow G12 8QQ, United Kingdom}

\author{Brian Lantz}
\affiliation{Stanford University, Stanford, CA, USA}

\author{Richard Mittleman}
\affiliation{LIGO Laboratory, Massachusetts Institute of Technology, Cambridge, MA 02138, USA}

\author{Hugh Radkins}
\affiliation{LIGO Hanford Observatory, Richland, WA 99352, USA}

\author{Bas Swinkels}
\affiliation{European Gravitational Observatory (EGO), I-56021 Cascina, Pisa, Italy}

\begin{abstract}

Gravitational wave observatories have always been affected by tele-seismic earthquakes leading to a decrease in duty cycle and coincident observation time. In this analysis, we leverage the power of machine learning algorithms and archival seismic data to predict the ground motion and the state of the gravitational wave interferometer during the event of an earthquake. We demonstrate improvement from a factor of 5 to a factor of 2.5  in scatter of the error in the predicted ground velocity over a previous model fitting based approach. The level of accuracy achieved with this scheme makes it possible to switch control configuration during periods of excessive ground motion thus preventing the interferometer from losing lock. To further assess the accuracy and utility of our approach, we use IRIS seismic network data and obtain similar levels of agreement between the estimates and the measured amplitudes. The performance indicates that such an archival or prediction scheme can be extended beyond the realm of gravitational wave detector sites for hazard-based early warning alerts.
\end{abstract}

\maketitle

\section{Introduction}
With the advent of gravitational wave (GW) astronomy, it is essential to maximize the duty cycle of second-generation gravitational-wave detectors such as the Laser Interferometer Gravitational-wave Observatory (LIGO) \cite{aligo}, Virgo \cite{avirgo}, and GEO600 \cite{Gr2010} detectors.
Any increase in duty cycle increases the sensitivity of GW searches, including the observations of binary black hole mergers and binary neutron stars \cite{AbEA2016a,AbEA2016e,AbEA2017a,AbEA2017c,abbott2017gw170608,AbEA2017b, ligo2018gwtc} . GWs from these induce small displacements in the detectors, which are designed to be free from environmental disturbances and limited only by processes of fundamental physics.
These detectors are subject to non-Gaussian noise transients due to either internal behavior of the instrument or interactions between the detector and its environment \cite{AbEA2016f}. To minimize the effect of the environment, the LIGO detectors contain 200,000 auxiliary channels which are designed to monitor both the behavior of the instrument and the environmental conditions.
A subset of these is physical environmental monitor sensors, including seismometers, magnetometers, microphones, and many others. Advanced LIGO \cite{aligo} and  Advanced Virgo \cite{avirgo} have in particular driven the development of both seismic \cite{BeCa2016} and rotation \cite{VeHa2014} sensors. Seismic sensors in particular are useful for measuring any source of ground motion that can couple into the interferometers. LIGO seismic isolation systems by means of passive and active isolation provide noise suppression above 0.1 Hz \cite{AbAd2002,StAb2009,MaLa2015} but are not effective against earthquake-related ground motion \cite{CoSt2015,CoEa2017}. The surface waves so produced hinder the process of keeping the instrument at a linear operating point and often induces higher frequency noise by up-converting low-frequency optical motion.

\begin{figure*}[!htb]
  \includegraphics[width=\textwidth]{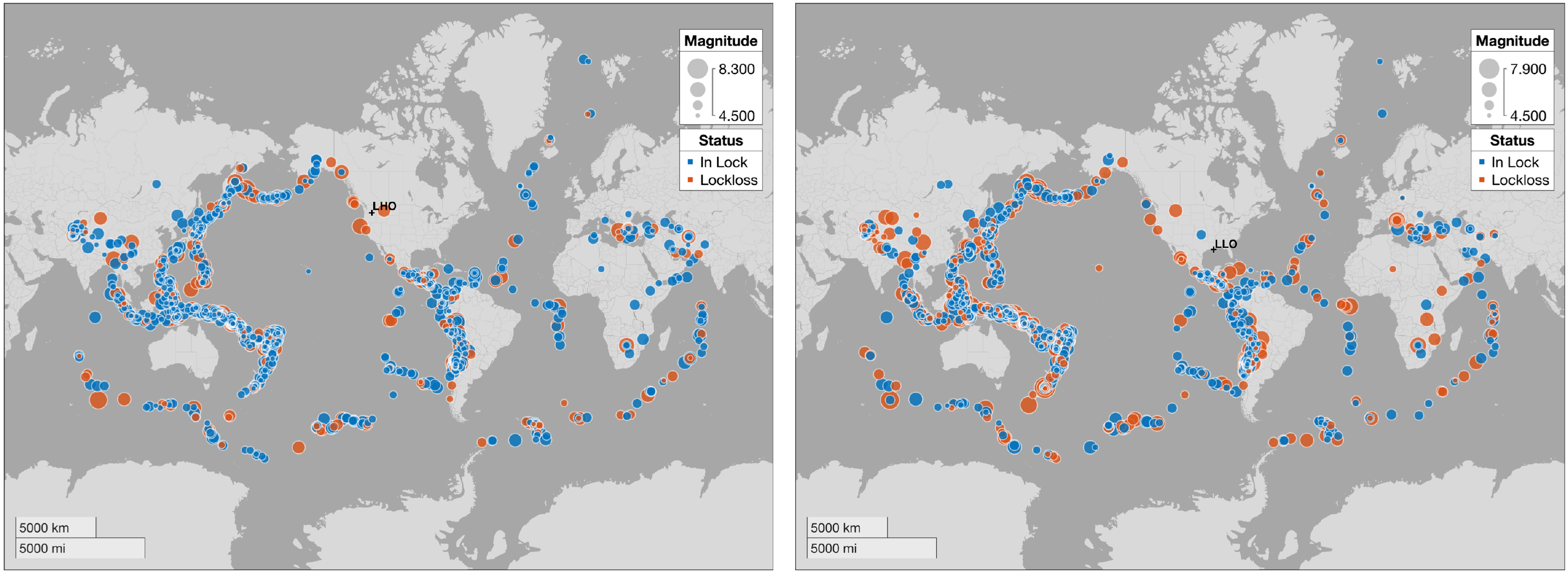}
 \caption{Impact of earthquakes happening worldwide on LIGO Interferometers at Hanford
and Livingston. Points marked in orange indicate the instances when the resulting ground the motion lead to a loss of control over the ability to keep the interferometer at its operating point resulting in a lockloss.}
 \label{fig:eq_worldwide}
\end{figure*}

\begin{figure*}[!htb]
  \includegraphics[width=\textwidth]{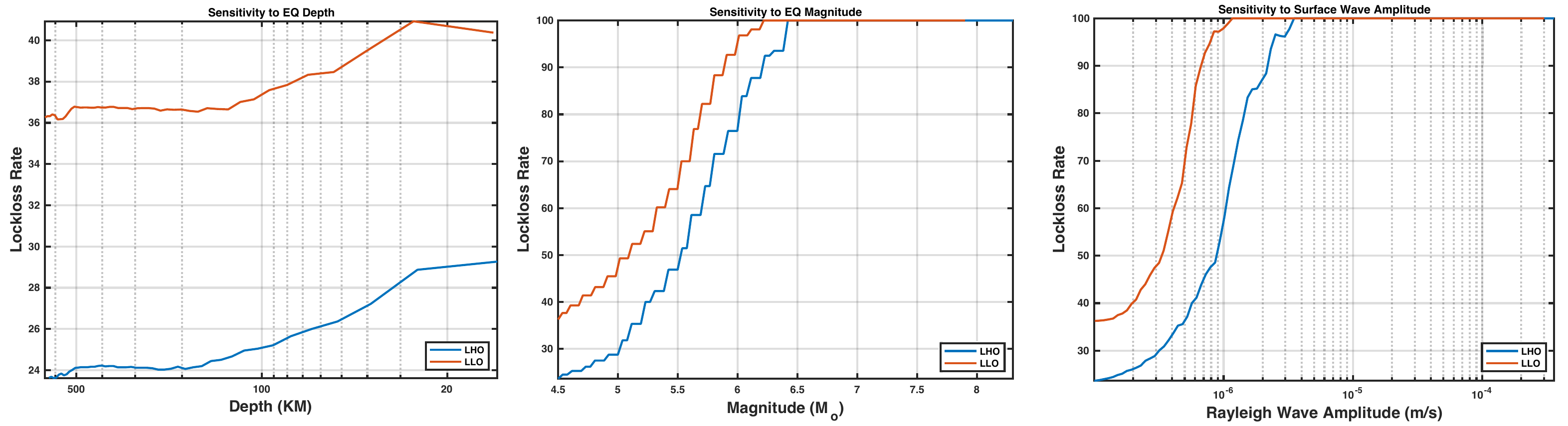}
 \caption{Plot shows the sensitivity of both the LIGO detectors to earthquakes in terms of distance to the hypocenter from the surface, earthquake magnitude and the Rayleigh wave amplitude.}
 \label{fig:lockloss_sensitivity}
\end{figure*}

Earthquake Early Warning (EEW) is a burgeoning field dedicated to the rapid detection and characterization of earthquakes as well as the dissemination of that information to people and infrastructure in their path \cite{Al2012,KuAl2013a,KuAl2013b,KuHe2014,CoLa2009a,CoLa2009b,BoAl2014,HoKa2008,HoEA2011c,StAl2016}.
Worldwide, many seismic and geodetic (GPS) sensor arrays exist that produce rapid earthquake information products, from magnitude and location estimates to regional centroid moment tensors (CMTs) and advanced slip inversions.
With wide-ranging public warning systems in Mexico and Japan and smaller-scale systems in many other countries, warnings from seconds to minutes are now available to reduce the impact of earthquakes on society \cite{StAl2016}.
The short warning times arise out of the physical processes that drive the earthquake rupture, where the warning is given by seismometers measuring P-waves ($\approx$\,8\,km/s) and S-waves ($\approx$\,4\,km/s).
Reliability of these estimates are one of the most important aspects of EEW systems. Their improvements generally rely on increasing the number of stations involved in the warning decisions as well as increasing alarm thresholds on ground motion, both seeking to limit the number of false positives \cite{KuCo2015}. Both of these strategies come at the cost of decreasing the warning time. 
As these systems minimize the time required to calculate the source parameters of earthquakes (i.e., their location and magnitude), it becomes important to predict with high accuracy the ground motion that the earthquakes will cause as a function of location and distance.

This paper is organized as follows: Section~\ref{sec:motivation} talks about the sensitivity of the GW interferometers to earthquakes and the previous attempts to model them. In Section~\ref{sec:data} we describe how the seismic data was obtained at the site as well as the IRIS seismic array \cite{IRIS,trabant2012data}. Section~\ref{sec:MLA} describes the deployed regression and clustering techniques and their requirements. Finally, the relative performance of various prediction algorithms along with the ability to guess the state of the interferometer are covered in Section~\ref{sec:results}.

\section{Impact Assessment}\label{sec:motivation}
Figure~\ref{fig:eq_worldwide} depicts the distribution of global seismic events and their respective effect on the state of the GW interferometers during LIGO's first and second observation run. The orange circles (scaled as per the magnitude) represent scenarios where the ground motion was high enough to cause instabilities leading to a loss of resonance in the cavity (lockloss). Sensitivities to parameters such as magnitude and surface wave amplitude are shown in Figure~\ref{fig:lockloss_sensitivity} where the steepness of the curve indicates higher sensitivity to the respective parameter. As expected, LLO is seen to be more vulnerable to ground shaking which can be attributed to its local geology and soil properties \cite{Daw2004}. The primary goal of LIGO/Virgo EEW methods would be to generate reliable relations between earthquake source parameters and ground motion metrics. Examples in the time domain include peak ground acceleration, peak ground velocity, and peak
ground displacement, while in the frequency domain there are spectral accelerations, velocities, displacements and predominant periods \cite{Do2003}.  Early estimates of magnitudes tend to underestimate the energy released from the tectonic motion and hence the early estimates of ground velocity amplitudes often tend to be not as accurate as the later values. The effects of these errors are particularly pronounced for larger earthquakes, where the estimates of the fault lengths become more important.
Thus, these larger earthquakes tend to have their amplitudes under-predicted.
The loss of performance that results from use of the rapid estimates is acceptable for the purpose of issuing rapid warnings. Direct displacement measurement based on non-inertial GPS instruments has been shown to generate rapid magnitude estimates within the first minute of rupture and in many cases even before rupture is complete \cite{MeCr2015}.

In previous work, we used advances in early earthquake warning to develop a low-latency earthquake early warning client named \emph{Seismon} \cite{CoEa2017}. This system uses a real-time event messaging system of the U.S. Geological Survey (USGS) to mitigate the effects of tele-seismic events on ground-based gravitational-wave detectors. 
Using information about the earthquake source characteristics such as magnitude, depth, and distance from the site, ground motion velocity at the site was predicted.
In the initial version of the algorithm, we used an empirical fit to an equation derived to account for physical effects. This equation succeeded in predicting peak ground velocity such that 90\% of events had a measured ground velocity within a factor of five of the predicted value. Although the fit was derived with physical effects in mind, it was predominantly an empirical construction. To make accurate assessments of whether the gravitational-wave detectors will be affected, we prefer the relative error in the ground velocity predictions, $\rm Rf_{amp}$, to be close to a factor of two, which is much smaller than the factor of five scatter seen. These predictions have two purposes. Firstly, they provide a meaningful metric which on-site-staff at the detectors can use to plan the response to the incoming earthquake. The response could be in the form of switching seismic isolation loops to steer the interferometer to a more robust configuration keeping it locked although with a lesser sensitivity \cite{BiWa2018}. The predictions also serve as inputs to the algorithms that make predictions on upcoming downtime, which could be utilized to perform opportunistic maintenance to rectify problems typically scheduled for weekly maintenance periods.

 \begin{figure}[!htb]
 \includegraphics[scale=0.7]{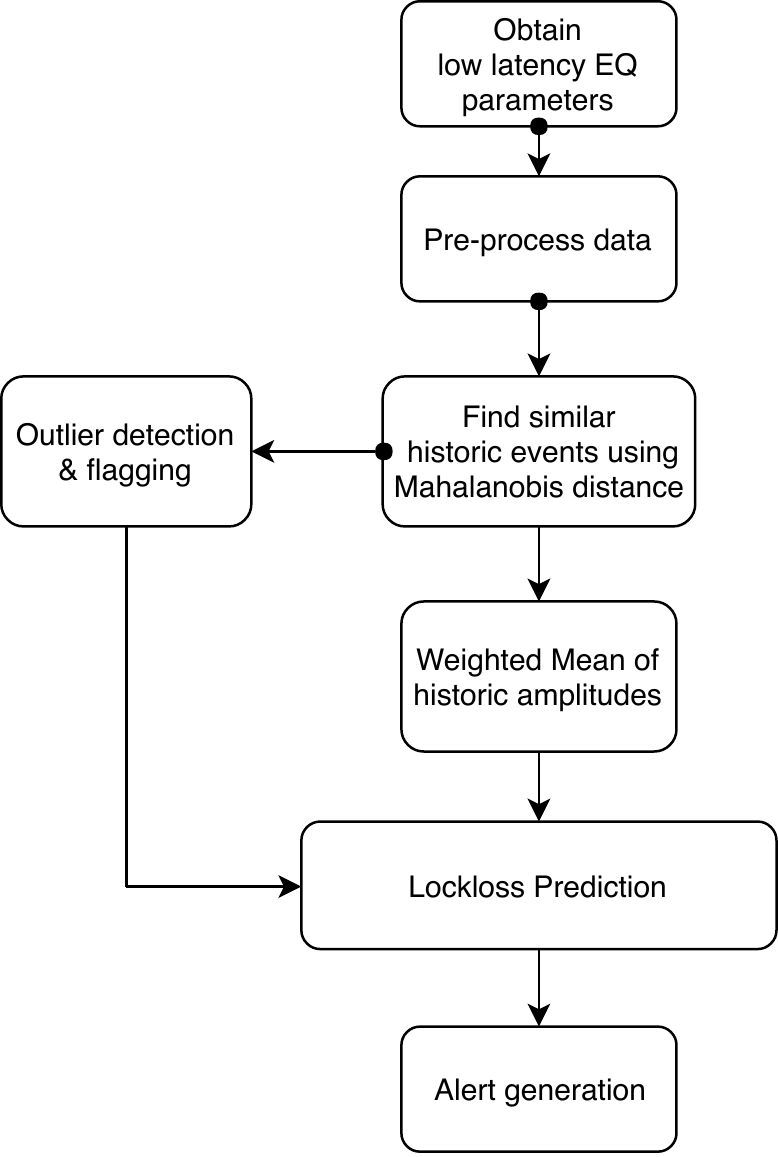}
 \caption{Process flowchart depicting the low latency earthquake warning pipeline.}
 \label{fig:flowchart}
\end{figure}   

    \begin{table*}[!htb]
        \centering
            \renewcommand\arraystretch{1.2}         
            \caption[ML Performance Table]{Rf amplitude prediction performance for different ML algorithms. The top and bottom panel give the percentage of events correctly predicted within a factor of 2.5 by each of the ML algorithm for simulated and real data.}
            \begin{tabular}{|*{5}{c|}}\cline{2-5}
                \multicolumn{1}{c|}{}& Deep Neural Nets & Stacked Ensemble & GPR & Clustering \\
                \hline              
                $LIGO \; Livingston $ & \diagbox[]{$85 \%$}{$89 \%$}& \diagbox[]{$89 \%$}{$93 \%$} &\diagbox[]{$87 \%$}{$94\%$}&\diagbox[]{$94\%$}{$98\%$}\\ 
                \hline              
                $LIGO \; Hanford $ & \diagbox[]{$84 \%$}{$86 \%$}& \diagbox[]{$88 \%$}{$91 \%$} &\diagbox[]{$89 \%$}{$92\%$}&\diagbox[]{$92\%$}{$97\%$}\\          
                \hline
            \end{tabular}
            \label{table:comparitive_study}
    \end{table*} 
    
\section{Methods: Seismic data.}\label{sec:data}

The first part of our analysis uses data obtained from the GW sites. For each earthquake event from the archival database, we take the vertical component of broadband data which is converted to ground velocity (m/s) using a frequency dependent calibration factor appropriate for each seismometer. Time-series are chosen to encompass the P-wave arrival and surface wave calculated assuming a (very conservative) seismic velocity of 2\, km/s \cite{shearer2009introduction}. As the seismometers located at the end and center stations observe similar values for the relevant frequency band, we use the center station sensor for rest of our analysis.

\begin{figure}[!htb]
 \includegraphics[width=\linewidth]{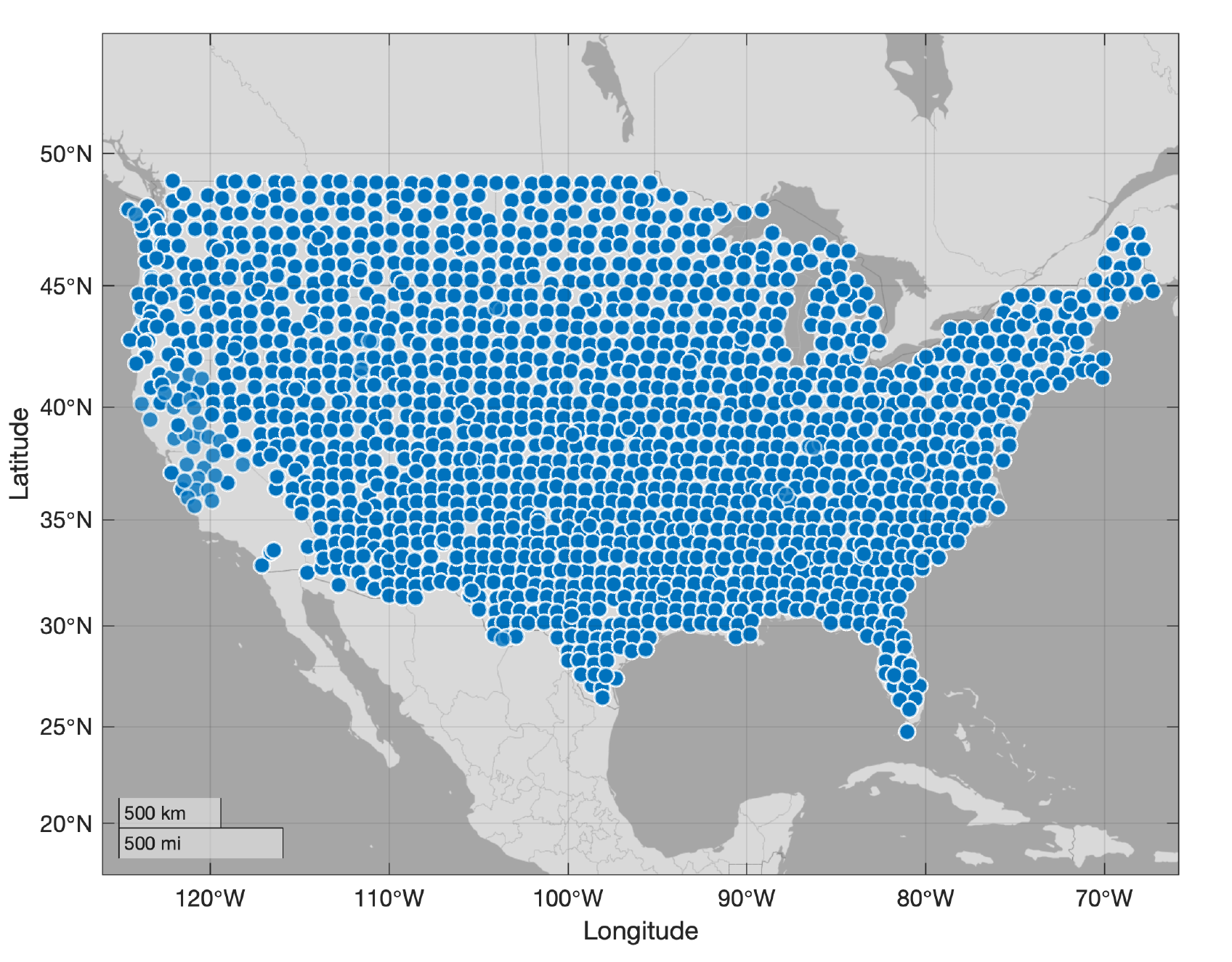}
 \caption{Locations of IRIS seismometer array used to collect the earthquake data from January 2006 to November 2017.}
 \label{fig:IRIS_Array}
\end{figure}

We also perform an analysis of seismic time-series that were made available through IRIS seismometer array, covering the last ten years. These stations covering all the conterminous US states (see Figure \ref{fig:IRIS_Array}) have time-series with response between 10\,mHz to 10\,Hz.  A variety of sources including anthropogenic and atmospheric disturbances contribute to the observed seismic signal. One source present across the world is the oceanic microseism around 0.3\,Hz that dominate seismic ground spectra everywhere on Earth \cite{HMS1963,ToLa1968,Ces1994,FKK1998}. Systematic processing of IRIS data from January 2006 to November 2017 lead us to create a database of around 733208 earthquakes. The analysis was restricted to earthquakes with magnitude ranging from 6.0 to 9.1, which covers the range that is likely to significantly effect the gravitational-wave detectors. In Section~\ref{sec:results} we report the performance of the prediction scheme in the presence of variation in local seismic spectra arising from affects such as local geology, topography and proximity to urban settlements.

\section{Methods: MLA description} \label{sec:MLA}

Machine Learning (ML) has recently become an important aspect of EEW and seismology in general. For example,
the \emph{MyShake} EEW system uses artificial neural networks to differentiate earthquake and human motions, with 98\% of earthquake records within 10\, km of the epicenter correctly identified, and only 7\% of people-induced transients appearing to be earthquakes to the algorithm  \cite{KoAl2016}.
They have been used to differentiate earthquakes from other seismic transients \cite{KuYi2011,KoUs2016,PeGh2017}, discriminate between deep and shallow micro-earthquakes \cite{MoHo2016} and to add to undersampled or missing traces \cite{JiMa2017}. In addition, they have been used to make full-wave tomography images \cite{DiLe2011}. The idea of our analysis is to compare historical ground velocity measurements to predictions made using different machine learning algorithm techniques. The parameters that enter the predictions are $M$, the magnitude of the earthquake, $h$, the depth, $log(r)$, the distance to the detectors in log scale, $\theta$ and $\phi$, the latitude and longitude, and $\alpha$, the earthquake azimuth relative to the detector. On longer timescales, the earthquake slip inversion, strike, rake, and dip, and the moment tensor values, $M_{rt}$, $M_{tp}$, $M_{rp}$, $M_{tt}$, $M_{rr}$, and $M_{pp}$ are also available as additional parameters which could be used in the future to improve the prediction accuracy, albeit not at low latency. The target variables correspond to peak ground velocities measured using seismometers. This scheme is advantageous over the analytical equation in a few ways. First of all, by switching to ML, we eliminate the dependence on a functional form.
Second, it trivially includes more parameters, such as latitude, longitude, and earthquake azimuth relative to the detector above and beyond the initial analytical formalism.

\begin{figure*}[!htb]
\hspace*{-0.5cm}
 \includegraphics[width=\textwidth]{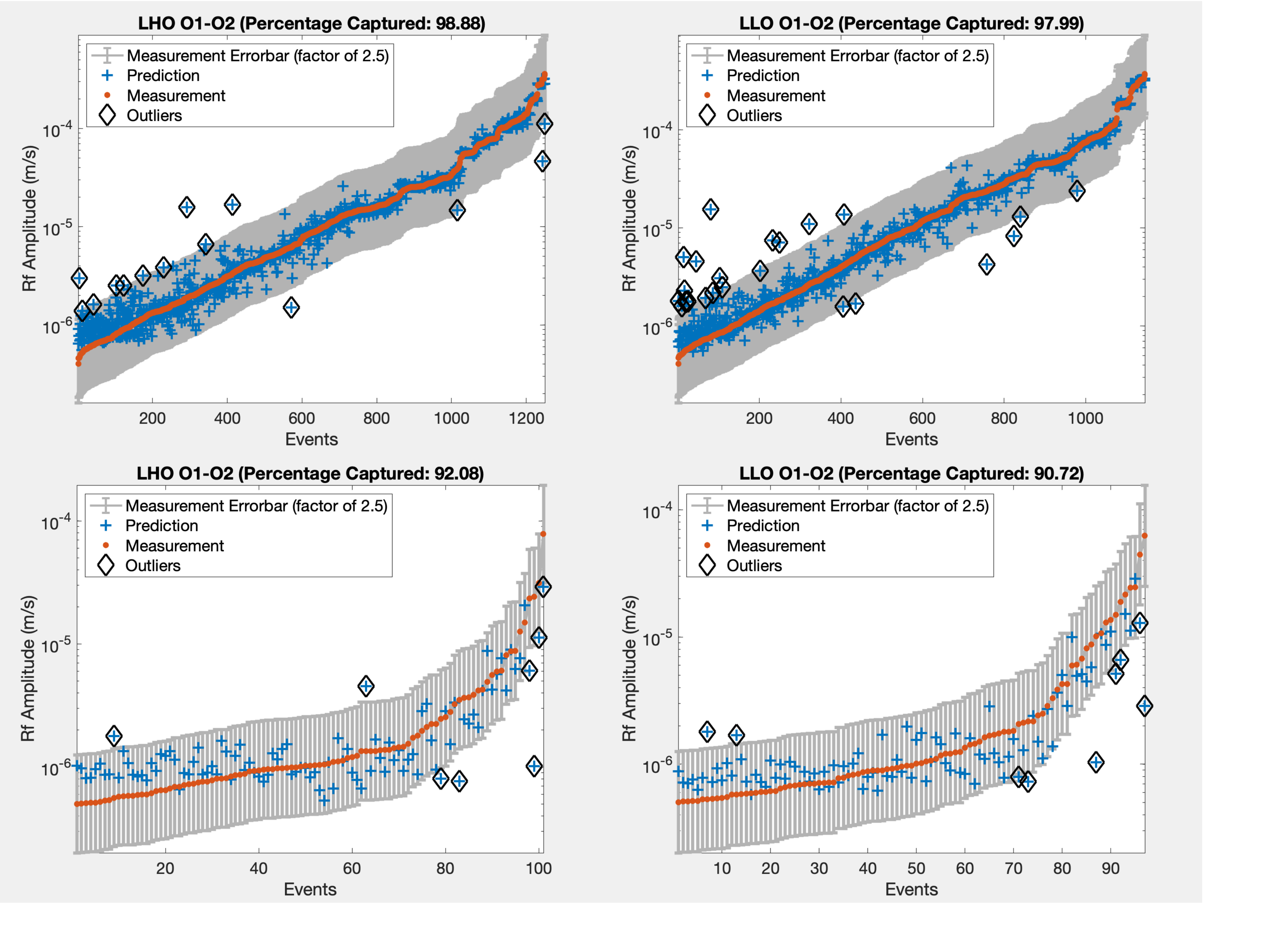}
 \caption{Fit of peak velocities seen during O1-O2 at the interferometers (LHO, LLO) using Mahalanobis distance-based clustering. Results on simulated and real data are respectively shown in the top and bottom rows. The events have been ordered by their measured peak ground velocity (in orange), and grey error bar corresponds to a factor of 2.5 within the measured value. More than 90\% of predictions (in blue) are within a factor of 2.5 of the measured value.}
 \label{fig:regression_LIGO}
\end{figure*}

\begin{figure*}[!htb]
\hspace*{-0.5cm}
 \includegraphics[width=\textwidth]{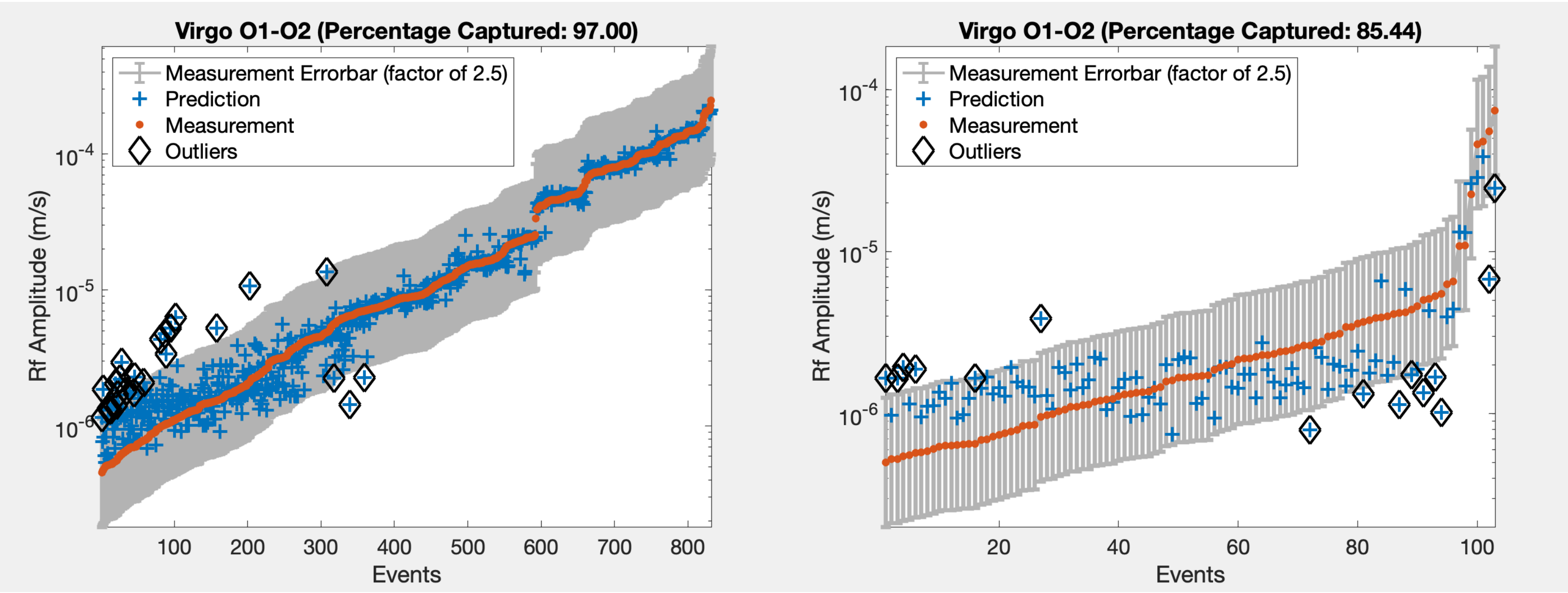}
 \caption{Fit of peak velocities seen during O1-O2 at the Virgo Interferometer}
 \label{fig:regression_Virgo}
\end{figure*}

In particular, we compare the efficiency of two different machine learning approaches: regression and clustering. Within regression, we evaluate the performance of the Tensorflow implementation of deep neural networks (DNN) \citep{Abadi:2016:TSL:3026877.3026899}, stacked ensemble regressors \citep{wolpert1992stacked} and Gaussian Process Regression (GPR) \citep{rasmussen2006gaussian}, while in clustering we use a Mahalanobis distance \citep{Mahalanobis} based similarity search to make the predictions. The performance of each algorithm is accessed using real and simulated datasets. For each dataset, 80\% is used for training with the remaining 20\% used for predictions. As for simulated data, new samples are generated from each of the original datasets by creating a Gaussian jitter distribution centered around the parameter value followed by a random draw of samples from these distributions. Artificially adding noise (or jitter) to the predictor and response variables in a controlled fashion helps to improve the learning and prevent early stopping. The presence of noise enhances the ability of the ML algorithms to better learn and generalize to the underlying smooth, non-linear function. Variance of the jitter distribution should be chosen such that the synthetic samples created are neither  completely nonsensical nor very much identical to the original data set \cite{holmstrom1992using}. In most of the cases, USGS do provide us with the level of uncertainty associated with the estimated earthquake parameters. We set the variance of the jitter distribution such that the new samples generated around the original values are well within the measurement uncertainties of each of the variables. For a given input taken from the jittered population, the predicted output value would be the weighted average of the outputs of the original training samples with the weights proportional to the densities of the jittered distribution. For a very large number of such samples, the average predicted value turns out to be the Nadaraya-Watson kernel regression estimator \cite{nadaraya1964estimating} where the variance in jitter corresponds to the bandwidth of the estimator \cite{scott2015multivariate}. 

The deep neural network (DNN) that we employ to carry out the nonlinear regression has a topology inspired from generalized regression neural networks \cite{specht1991general}, but we back-propagate the errors and update the weights by training it through several epochs.   DNNs, in general, require larger data sets to learn the underlying function without overfitting the data and tend to be sensitive to the network architecture and the activation functions. We use a sequential network with nine dense layers with exponential linear unit activation and the first-order gradient-based Adam optimizer \cite{kingma2014adam}. In recent times, stacked ensemble regressors have also gained much prominence and are seen to consistently outperform other  competing algorithms in several datasets hosted at the Kaggle challenge \cite{Kaggle}. The first level consists of a set of base learners that are individually trained and cross-validated. Their predictions form input to a second level meta-learner regressor which is further trained to generate the final ensemble prediction.  Such systems are theoretically guaranteed to be the optimal learners in an asymptotic sense \cite{van2007super}. Success with DNN and ensemble techniques crucially depends on the amount of training data and could be sensitive to the hyper-parameters. As for the GPR, we optimize the respective hyper-parameters using Bayesian optimization and use a squared exponential kernel for covariance estimation. But the method scales as O($n^3$) thus resulting in high memory requirements and training time for large data sets.

Mahalanobis distance is the multi-dimensional generalization of z-score which tells you how many sigmas the data is away from the mean distribution. It is observed to be a very robust technique as it takes into account the covariances between the variables. Our clustering technique makes use of this metric to find the closest matching earthquakes that happened in the past. This scheme naturally lets one identify outlier earthquakes with no similar events in the archival data. Table~\ref{table:comparitive_study} compares the performance of the different MLAs. We find the clustering-based predictor to be the most accurate and select is as the default algorithm  for EEW pipeline shown in Figure~\ref{fig:flowchart}. In addition to having the best prediction accuracy, it has the following advantages.  Firstly, as there is no training involved, the need for hyper-parameter tuning is eliminated. As we are constantly monitoring the seismic data and appending the earthquake database, with time we expect a decrease in prediction error along with a reduction in the number of outliers.  

\section{Results}\label{sec:results}

\begin{figure*}[!htb]
 \includegraphics[width=\textwidth]{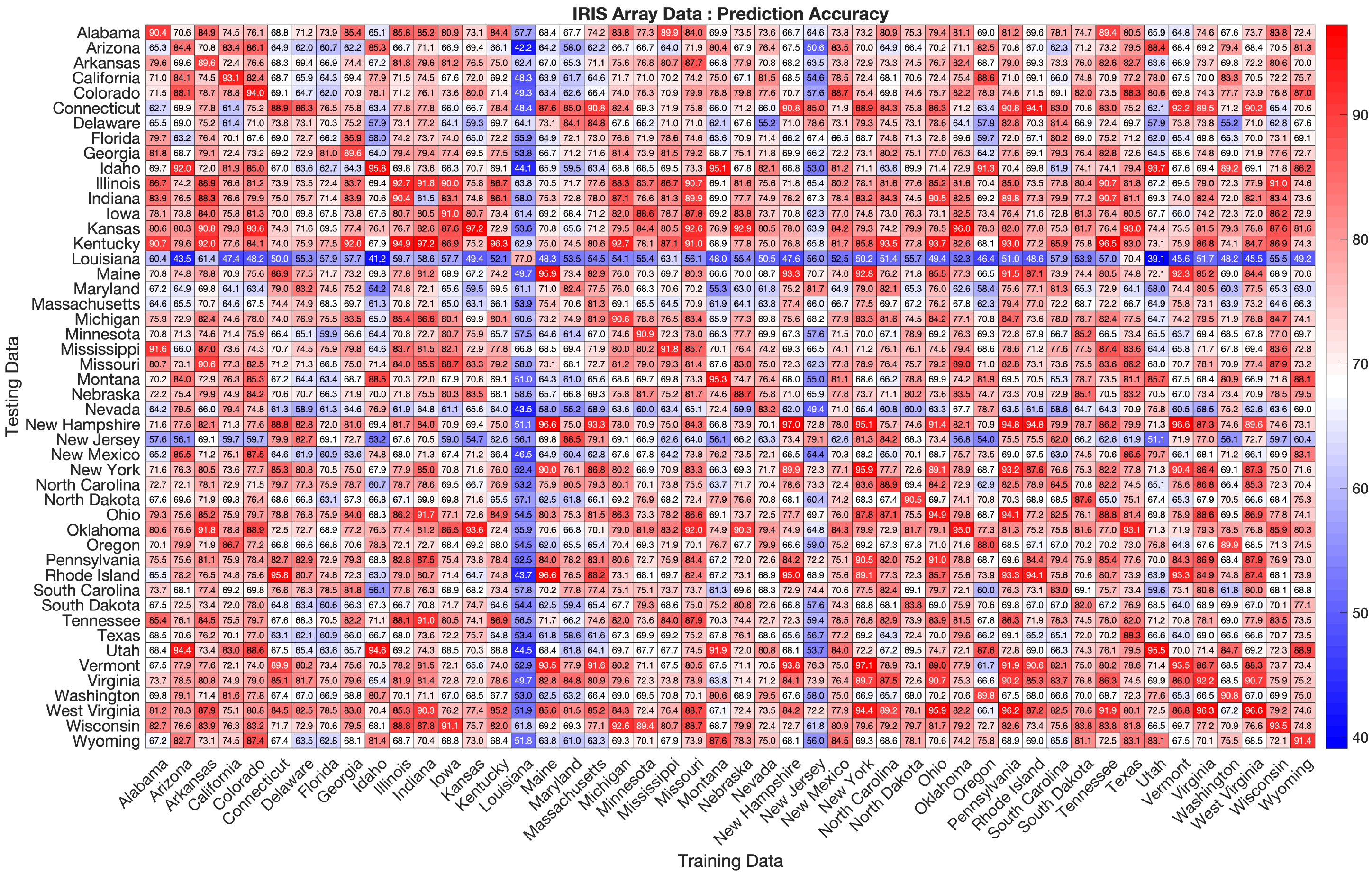}
 \caption{The heat-map on the right shows the ground motion prediction accuracy within each state making use of archival data from all the conterminous US states.}
 \label{fig:map}
\end{figure*}

Figures~\ref{fig:regression_LIGO} and~\ref{fig:regression_Virgo} show the prediction results from applying Mahalanobis based clustering on the simulated and real earthquake data. The events have been ordered by their measured peak ground velocity (in orange), and the error bar (in grey) corresponds to a factor of 2.5 within the measured value. As compared to the previously used analytic expression, this machine learning based scheme leads to a performance improvement from a factor of 5 to 2.5 in scatter of the error in the predicted ground velocity. For both real and simulated cases, accuracy in predictions (in blue) is observed to be above 90\% for LIGO and 85\% for Virgo datasets.We can attribute the improvement to the increased availability of data, the inclusion of more earthquake parameters and the usage of robust algorithms. This level of accuracy is sufficient to switch control configuration during periods of excessive ground motion thus preventing the interferometer from losing lock. As for the outliers, they don't seem to show any specific dependence concerning the input parameters. One reason could be that their parameter combination is rather uncommon, so the predicted amplitude is averaged across not-so-similar events. The general trend among outliers seems to be that the higher amplitude events are underestimated, and the lower amplitude events are over-estimated. Such outliers should decrease as we gather more training data. Bringing down the mismatch, especially at the lower amplitude side, would require improvements in signal to noise ratio of the ground velocity measurements carried out at the site along with the need for more detailed initial earthquake information. Availability of low latency information regarding the centroid moment tensors associated with the earthquake could be one such solution.

\begin{table*}[!htb]
    \centering
    \begin{minipage}{.45\textwidth}

        \centering

 \begin{tabular}{  | l | c | c | r   } 
 \hline 
\textbf{LHO} & In Lock & Lockloss  \\ 
\hline 
True Positives           & 39 & 10  \\ 
False Positives            & 1 & 3 \\ 
  \hline 
Precision      & 0.97 & 0.77 \\  
Sensitivity     & 0.93& 0.91 \\   
Specificity    & 0.91 & 0.93 \\   
\hline                 
\end{tabular} 
 
    \end{minipage}%
    \begin{minipage}{0.45\textwidth}
        \centering

\begin{tabular}{  | l | c | c | r   } 
 \hline 
\textbf{LLO} & In Lock & Lockloss  \\ 
\hline 
True Positives           & 49 & 16  \\ 
False Positives           & 3 & 2 \\ 
   \hline 
Precision      & 0.94 & 0.89 \\  
Sensitivity      & 0.96& 0.84 \\   
Specificity    & 0.84 & 0.96 \\   
\hline                 
\end{tabular}

    \end{minipage}
\caption{Performance analysis of lockloss prediction models for LHO and LLO. Each of them respectively has an accuracy of 92\% and 93\%.} 
\label{table:lockloss_prediction}  
\end{table*}

We also demonstrate the resourcefulness of the above scheme by making predictions across the United States using the data recorded by the IRIS network. For ground motion recorded in each state, we perform the similarity search using archival data  and compare the prediction accuracy as shown in Figure~\ref{fig:map}. For each row representing a state, we show how well we can make ground velocity predictions based on archival events from itself as well as other states. For at least 24 states, the accuracy is seen to be above 90\% when we use its training data for predicting the new events. The variation seen in predictability along the diagonal might be due to the differences in local geology across the US. This observation of unpredictability could be beneficial for future site selection surveys looking for suitable locations for next-generation interferometers. High values seen along several of the off-diagonal terms are due to the nearly identical response to tele-seismic events could mean the corresponding similarity within their geological properties such as shear velocity profiles, elasticity, and local soil density, etc. 
In future, state-wise seismic modeling could use this information to augment the state-wise data with the ones from similar states especially if the original dataset is sparse. The performance shows that the archival data-based prediction scheme can be extended beyond the realm of gravitational-wave detector sites for hazard-based early warning alerts. For the state of Louisiana, the data from previous states do not seem to make predictions with a high level of accuracy. As expected, comparatively better performance is obtained from the neighboring states of Texas, Arkansas, and Mississippi. Since similar anomalies are not observed in Florida and Texas data, proximity to the Gulf of Mexico does not seem to a strong reason for this behavior. Usage of multiple sensors within each state and the adoption of standardized data analysis procedure which includes the individual sensor calibration and data preprocessing minimizes the likelihood of systematic and instrumental effects to bias the results. Another possibility is from peculiarities in local geology, but further studies would be required to understand the exact cause of this anomaly. 

The main benefit of ground velocity predictions for gravitational-wave detectors is to inform predictions of whether an earthquake will cause the loss of data for the detector. We have previously developed techniques for preventing earthquakes from causing the loss of data taking if advanced notice is given \cite{BiWa2018}. In this work, we  use the previously described clustering technique to develop a lockloss prediction model as well. 
We use the same set of inputs to the algorithm as in the ground velocity prediction case, but also include the ground velocity predictions themselves as inputs. To generate the target variable, we take times when the gravitational-wave detectors lost the ability to take data during an earthquake and assign a value of 1, and a 0 otherwise. Acknowledging that there is a trade-off between false-alarm probability and efficiency standard, 
we can make predictions for the inliers with an accuracy above 92\,\%, keeping the associated false-alarm probability to be less than 10\,\% (See table~\ref{table:lockloss_prediction} ). Given this scenario of an impending lockloss, it would be more desirable to switch to a configuration that provides enough freedom for common mode motion but at the same time enforces the best possible suppression for the local differential motion.

\section{Conclusions}
In conclusion, we have used MLAs to predict peak ground velocities from tele-seismic earthquakes. The estimated ground velocity is used to forecast the potential effect of earthquakes on gravitational-wave detectors and issue near real-time alerts at the site. The alert system based on this scheme has been implemented at the Advanced GW observatories and will be used shortly to switch seismic filters at very low frequencies. Given the significant interest in accurate ground velocity predictions for EEW systems in general, we believe the techniques here are beneficial beyond the gravitational-wave community. While we focus on the prediction of peak ground velocity here, in future we will explore the possibility of using historical seismometer data along with CMT parameters to predict radiation patterns associated with the fault rupture. This would provide a way to directly measure transfer functions between ground motion very near the earthquake source and those in areas of significant seismological hazard, such as in the Los Angeles basin.

\textbf{Code availability.}
The code to reproduce the analysis is open-source and available at https://github.com/ligovirgo/seismon/ for public download.

\textbf{Acknowledgments.}
NM acknowledges Council for Scientific and Industrial Research (CSIR), India, for providing financial support as Senior Research Fellow.  MC was supported by the David and Ellen Lee Postdoctoral Fellowship at the California Institute of Technology. Authors express thanks to Duncan Agnew, Rich Ormiston and Brian O'Reilly for their valuable comments and suggestions. LIGO was constructed by the California Institute of Technology and Massachusetts Institute of Technology with funding from the National Science Foundation and operates under cooperative agreement PHY-0757058. This paper has been assigned LIGO document number LIGO-P1800312. Global Seismographic Network (GSN) is a cooperative scientific facility operated jointly by the Incorporated Research Institutions for Seismology (IRIS), the United States Geological Survey (USGS), and the National Science Foundation (NSF), under Cooperative Agreement EAR-1261681.
The facilities of IRIS Data Services and specifically the IRIS Data Management Center were used for access to waveforms, related metadata, and derived products used in this study. IRIS Data Services are funded through the Seismological Facilities for the Advancement of Geoscience and EarthScope (SAGE) Proposal of the National Science Foundation under Cooperative Agreement EAR-126168

\bibliographystyle{iopart-num}
\bibliography{references}

\end{document}